\documentclass[
final            % use final for the camera ready runs
,numberedheadings % uncomment this option for numbered sections
  ]
  {aipproc}

\layoutstyle{8x11single}

\usepackage{axodraw,color}

\newcommand{\lapprox}{%
\mathrel{%
\setbox0=\hbox{$<$}
\raise0.6ex\copy0\kern-\wd0
\lower0.65ex\hbox{$\sim$}
}}
\newcommand{\gapprox}{%
\mathrel{%
\setbox0=\hbox{$>$}
\raise0.6ex\copy0\kern-\wd0
\lower0.65ex\hbox{$\sim$}
}}

\begin{document}

\title{Isospin breaking in pion and $K_{e4}$ form factors}


\author{V. Bernard}{
  address={Groupe de Physique Th\'eorique, Institut de Physique Nucl\'eaire\\
B\^at. 100, CNRS/IN2P3/Univ. Paris-Sud 11 (UMR 8608), 91405 Orsay Cedex, France}
}

\author{S. Descotes-Genon,}{
  address={Laboratoire de Physique Th\'eorique, CNRS/Univ. Paris-Sud 11 
(UMR 8627), 91405 Orsay Cedex, France}
}

\author{M. Knecht\footnote{Speaker}~}{
  address={Centre de Physique Th\'eorique,
 CNRS/Aix-Marseille Univ./Univ. du Sud Toulon-Var (UMR 7332)\\
CNRS-Luminy Case 907, 13288 Marseille Cedex 9, France}
}

\begin{abstract}
 Isospin breaking in the $K_{\ell 4}$  form factors induced by the difference between charged and
neutral pion masses is discussed within a framework built on suitably subtracted dispersion representations. 
The $K_{\ell 4}$ form factors are constructed in an iterative way up to two loops in the low-energy expansion 
by implementing analyticity, crossing, and unitarity due to two-meson intermediate states. Analytical expressions 
for the phases of the two-loop form factors of the $K^\pm\to\pi^+\pi^- e^\pm \nu_e$ channel are presented, allowing 
one to connect the difference of form-factor phase shifts measured experimentally (out of the isospin limit) and the 
difference of $S$- and $P$-wave $\pi\pi$ phase shifts studied theoretically (in the isospin limit). The dependence 
with respect to the two $S$-wave scattering lengths $a_0^0$ and $a_0^2$ in the isospin limit is worked out in a general way, 
in contrast to previous analyses based on one-loop chiral perturbation theory. The results on the phases of the 
$K^\pm\to\pi^+\pi^- e^\pm \nu_e$ form factors obtained by the NA48/2 collaboration at the CERN SPS are reanalysed 
including isospin-breaking correction to extract values for the scattering lengths $a_0^0$ and $a_0^2$.
\end{abstract}

\maketitle

%%%%%%%%%%%%%%%%%%%%%%%%%%%%%%%%%%%%%%%%%%%%
%% MAINMATTER
%%%%%%%%%%%%%%%%%%%%%%%%%%%%%%%%%%%%%%%%%%%%

\section{Introduction}

Very accurate information on the $\pi\pi$ S-wave scattering lengths in the isospin limit $a_0^0$ and
$a_0^2$ is now available from several experimental processes, $K^\pm\to\pi^0\pi^0\pi^\pm$ \cite{NA48-Kpi3},
pionic atoms \cite{Adeva:2011tc}, and the $K_{e4}$ semi-leptonic decay $K^\pm \to \pi^+ \pi^- e^\pm \stackrel{_{(-)}}{\nu_e}$.
In the last case, progress has been particularly impressive in recent years. The NA48/2 Collaboration
\cite{Batley:2007zz,Batley:2010zza}  at the CERN SPS has collected $\sim 1\,100\,000$ events. This represents 
more than twice the statistics obtained by the previous experiment at the Brookhaven AGS, where the BNL-E865 collaboration 
\cite{Pislak:2001bf,Pislak:2003sv} had collected $\sim 400\,000$ events, and an improvement by a factor of 
more than $35$ with respect to the Geneva-Saclay experiment \cite{Rosselet:1976pu}, the first high-statistics experiment 
of this type, which, almost 40 years ago, had collected $\sim 30\,000$ events.

Standard angular analysis of the $K_{e4}$ decay amplitude \cite{Cabibbo:1965zz,Berends:1968zz} 
shows that information on $\pi\pi$ scattering is contained in the phases of the form factors
that describe it. The interference term that two of these form factors produce in the differential decay rate 
has a phase that is measurable, and  given by the difference $\delta_S (s) - \delta_P (s)$ between 
the phases of the $S$ and $P$ partial-wave 
projections of the $\pi\pi$ scattering amplitude, as a consequence of Watson's theorem \cite{Watson52}.
The values of the scattering lengths can then be extracted upon fitting the experimentally measured
phase difference with the corresponding solution $\delta^{S-P}_{\rm Roy} (s ; a_0^0 , a_0^2)$ of the Roy equations:
\begin{equation}
\left[ \delta_S ( s) - \delta_P (s) \right]_{\rm exp}
= \delta^{S-P}_{\rm Roy} (s ; a_0^0 , a_0^2)
.
\label{data1}
\end{equation}
The Roy equations \cite{Roy:1971tc} rely on fixed-$t$ dispersion relations (i.e. analyticity, unitarity, crossing,
the Froissard bound) for the $\pi\pi$ amplitudes, $\pi\pi$ data at higher energies $s\gapprox 800$ MeV,
and isospin symmetry. Numerical solutions for these equations exist and can be constructed for arbitrary values
of the scattering lengths $a_0^0$ and $a_0^2$ belonging to the so-called universal band,
see Ref. \cite{Ananthanarayan:2000ht} for details.
In the real world, isospin is not an exact symmetry. It is explicitly broken by
electromagnetic corrections, and by the small effects induced by the quark-mass difference
$m_u - m_d$. While radiative corrections are considered in the analysis performed
by the NA48/2 Collaboration \cite{Batley:2007zz,Batley:2010zza}, there remain
small isospin-breaking (IB) effects related to the difference of the masses of
charged ($M_\pi$) and neutral ($M_{\pi^0}$) pion, $M_\pi \neq M_{\pi^0}$.
As emphasized in \cite{GasserKaon07}, it is important to account for these effects
in extracting the values of the scattering lengths from the data, given the level
of precision achieved. The evaluation of the relevant IB correction to the $K_{e4}$
matrix element and differential decay rate were subsequently worked out at one
loop precision in the chiral expansion \cite{Colangelo:2008sm}. This allows one to replace
Eq. (\ref{data1}) by the more appropriate relation
\begin{equation}
\left[ \delta_S ( s) - \delta_P (s) \right]_{\rm exp}
= \delta^{S-P}_{\rm Roy} (s ; a_0^0 , a_0^2) + 
\delta_{\rm IB}^{\rm 1~loop} (s ;  (a_0^0)_{\rm CA}  , (a_0^2)_{\rm CA}  )
,
\label{data2}
\end{equation}
where $\delta_{\rm IB}^{\rm 1~loop} (s ;  (a_0^2)_{\rm CA}  , (a_0^2)_{\rm CA}  )$
denotes the correction factor to the phase difference computed in Ref. \cite{Colangelo:2008sm}. 
Before commenting on it, let us quote the values \cite{Batley:2010zza} obtained
from the fit using solutions of the Roy equations provided
by Refs. \cite{Ananthanarayan:2000ht,DescotesGenon:2001tn}, 
and the correction from Ref. \cite{Colangelo:2008sm} 
[we quote here the result from ``Model B'']
\begin{equation}
a_0^0
= 0.2220 (128)_{\mbox{\footnotesize stat}} (50)_{\mbox{\footnotesize syst }} (37)_{\mbox{\footnotesize th}}
\, ,\qquad
a^2_0
= -0.0432 (86)_{\mbox{\footnotesize stat}} (34)_{\mbox{\footnotesize syst}} (28)_{\mbox{\footnotesize th}}
\, .
\label{NA48-2_fit_Model_B}
\end{equation}
As already mentioned, the correction $\delta_{\rm IB}^{\rm 1~loop} (s ;  (a_0^2)_{\rm CA}  , (a_0^2)_{\rm CA}  )$
is evaluated at next-to-leading (one-loop) order only,
which raises the issue of the possible sensitivity of the analysis to higher order corrections,
given the high accuracy of the experimental data. In addition, the NLO correction
computed in chiral perturbation theory necessarily involves the scattering lengths
fixed at their tree level (current algebra) values \cite{Weinberg:1966kf},
$(a_0^0)_{\rm CA}= 7 M_\pi^2/32\pi F_\pi^2$ and $(a_0^2)_{\rm CA}= - M_\pi^2/16\pi F_\pi^2$. 
Actually, higher-order effects were estimated
in Ref. \cite{Colangelo:2008sm}, but from a NNLO calculation of the scalar form factor
of the pion. This estimate accounts for almost all the theory error in Eq. (\ref{NA48-2_fit_Model_B}).
The same drawbacks are shared by other studies devoted to IB in $K_{e4}$ decays \cite{cuplov04,Cuplov:2003bj,Stoffer:2013sfa}.
The situation is then that one extracts the scattering lengths from a fit to
solutions of the Roy equations, which depend parametrically on the scattering lengths,
after having applied IB corrections evaluated for fixed and predefined values
of $a_0^0$ and $a_0^2$!
This limitation may induce a bias in the extraction of the scattering 
lengths from data based on Eq. (\ref{data2}), and it is important to be able to
quantify this effect. It is therefore necessary to develop a computational framework of isospin-breaking corrections
in the phases of the form factors
where the values of the scattering lengths are not unnecessarily restricted from the outset.
The outcome of such a construction should result in the replacement of Eq. (\ref{data2}) by
\begin{eqnarray}
\left[ \delta_S ( s) - \delta_P (s) \right]_{\rm exp}
= \delta^{S-P}_{\rm Roy} (s ; a_0^0 , a_0^2) + 
\delta_{\rm IB} (s ;  a_0^2  , a_0^2  )
,
\label{data_full}
\end{eqnarray}
where $\delta_{\rm IB} (s ;  a_0^2  , a_0^2  )$ is evaluated
at least at NNLO, and where $a_0^0$ and $a_0^2$ appear as
free parameters. How this goal can be achieved 
will be described in the sequel. Further details may be found
in Refs. \cite{Bernard:2013faa} and \cite{DescotesGenon:2012gv}, 
on which the present report is based.

Before starting, let us illustrate the issue with a simple example, 
leaving aside, for the sake of demonstration, violations of isospin symmetry.
In the isospin limit, the one-loop expressions of the $K_{\ell 4}$ form factors are well documented in 
the literature \cite{Bijnens90,Bijnens94}, and one finds for one of the form
factors involved in the decay channel of interest, $K^+ \to \pi^+\pi^-\ell^+\nu_\ell$,
\begin{equation}
F^{\mbox{\tiny{$+-$}}} (s,t,u) = \frac{M_K}{\sqrt{2} F_\pi} \left[
1 + \cdots + \frac{2 s - M_\pi}{2 F_\pi^2} J_{\pi\pi}^r (s) + \cdots
\right],
\label{example_1}
\end{equation} 
where the ellipses stand for additional contributions that play no role in the present discussion,
$s$ denotes the square of the invariant mass of the dipion system, $F_\pi$ is the pion decay constant,
and  $J_{\pi\pi}^r$ is the renormalized one-loop two-point function. 
In this expression of the one-loop form factor, no dependence on the scattering lengths is visible, neither
in this term nor in the omitted ones. However, in the computation of the form factors, the actual expression
in terms of the low-energy constants of the $\chi$PT Lagrangian \cite{Gasser:1984gg} reads
\begin{equation}
F^{\mbox{\tiny{$+-$}}} (s,t,u) = \frac{M_K}{\sqrt{2} F_\pi} \left[
1 + \cdots + \frac{2 s - 2 {\widehat m} B_0}{2 F_0^2}  J_{\pi\pi}^r (s) + \cdots
\right], \qquad\qquad [{\widehat m} = (m_u + m_d)/2]
\label{example_2}
\end{equation} 
which agrees with the previous expression (\ref{example_1}) if the leading-order relations $F_\pi = F_0$ and 
$M_\pi^2 = 2 {\widehat m} B_0$ are used [this is the appropriate order to consider in this example], explaining 
why the expression  (\ref{example_1})  is usually quoted. 
However, it is not straightforward to reinterpret the expression (\ref{example_2}) in terms of
the $\pi\pi$ scattering lengths $a_0^0$ and $a_0^2$: they are both proportional
to $2 {\widehat m} B_0$ at lowest order \cite{Gasser:1984gg}, but there are infinitely many combinations of 
$M_\pi^2$, $a_0^0$, and $a_0^2$ that sum up to $2 {\widehat m} B_0$ at this order.
Even if a contribution from the $I=2$ channel is forbidden  by the 
$\Delta I = 1/2$ rule of the corresponding weak charged current, the question still remains how to determine the combination
that gives the correct dependence on $a_0^0$. Obviously, the information provided
by Eq.~(\ref{example_2}) alone does not allow for an unambiguous answer.
As can easily be guessed, the missing link is provided by unitarity. The function $J_{\pi\pi}^r$ encodes the discontinuity
of the form factor $F^{\mbox{\tiny{$+-$}}} (s,t,u)$ along the positive real $s$-axis, which involves the $I=0$ 
$\pi\pi$ partial wave in the channel with zero angular momentum as a final-state interaction effect~\cite{Bijnens94}. 
A careful analysis shows that, at one-loop order, 
Eqs.~(\ref{example_1}) and (\ref{example_2}) actually read
\begin{equation}
F^{\mbox{\tiny{$+-$}}} (s,t,u) = \frac{M_K}{\sqrt{2} F_\pi} \left[
1 + \cdots + \left( \frac{s - 4 M_\pi^2}{F_0^2} + 16 \pi a_0^0 \right) J_{\pi\pi}^r (s) + \cdots
\right].
\label{example_3}
\end{equation} 
Let us stress that, barring higher-order contributions presently not under discussion,
the three representations are strictly identical. However, if one considers the scattering lengths $a_0^0$ 
and $a_0^2$ as free variables that have to be adjusted 
from a fit to experimental data, only the third form is actually suitable. It is certainly conceivable
to use the existing one-loop expressions of $K^+_{e4}$ form factors, now including isospin-violating
effects \cite{cuplov04,Cuplov:2003bj}, and to repeat the above analysis for each separate 
contribution. But this would represent a rather cumbersome exercise,
and would anyway only give a result to one-loop precision. 
Instead, we will develop a more global approach, where the relevant unitarity 
properties are put forward explicitly from the start, and which, in addition,
holds at two-loop precision. This approach proceeds along the same lines
as those followed in order to establish the ``reconstruction theorem'' for
the $\pi\pi$ scattering amplitude (in the isospin limit) in Ref. \cite{Stern:1993rg}
and then implemented in order to construct an explicit two-loop representation
of this amplitude in Ref. \cite{Knecht:1995tr}.

\section{Two-loop representation of pion form factors with IB}

In order to dispense, at a first stage, with some of the kinematical complexities that beset
the discussion of the $K_{e4}$ form factors, we describe the  general method using the
simpler framework provided by the neutral and charged scalar form factors of the pion. 
These form factors are defined as [${\widehat m}\equiv (m_u + m_d)/2$]
\begin{eqnarray}
\langle \pi^0(p_1) \pi^0(p_2) \vert {\widehat m}({\overline u}u + {\overline d}d)(0)\vert\Omega\rangle
 &=&
 + F_S^{\pi^0}(s)
\quad { }
\nonumber\\
\langle \pi^+(p_{\mbox{\tiny{$ +$}}}) \pi^-(p_{\mbox{\tiny{$ -$}}}) \vert {\widehat m}({\overline u}u + {\overline d}d) (0)\vert\Omega\rangle
 &=&
- F_S^{\pi}(s) ,
\quad { }
\end{eqnarray}
and 
\begin{equation}
\frac{1}{2}\langle \pi^+ \pi^- \vert 
({\overline u}\gamma_\mu u - 
{\overline d}\gamma_\mu d)(0)\vert\Omega\rangle 
 =
(p_{\mbox{\tiny{$ -$}}}  - p_{\mbox{\tiny{$ +$}}})_\mu F_V^{\pi}(s) 
.
\end{equation}

The starting point of the construction is
provided by dispersive representations of the form factors 
and of the $\pi\pi$ scattering amplitudes. For the former, they write \cite{GasserMeissner91}
\begin{eqnarray}
F_S^{\pi^0}\!(s) &=& F_S^{\pi^0}\!(0) \! \left[
1 + \frac{1}{6}\langle r^2\rangle_S^{\pi^0}s + c_S^{\pi^0} \! s^2
+ U_S^{\pi^0}\!(s) 
\right]
\nonumber\\
F_S^{\pi}(s) &=& F_S^{\pi}(0) \! \left[
1 + \frac{1}{6}\langle r^2\rangle_S^{\pi}\,s + c_S^{\pi} \, s^2
+ U_S^{\pi}(s)
\right]
\nonumber\\
F_V^{\pi}(s) &=& 
1 + \frac{1}{6}\langle r^2\rangle_V^{\pi}\,s + c_V^{\pi} s^2
+ U_V^{\pi}(s)
,
\end{eqnarray}
with
\begin{eqnarray}
U_S^{\pi^0}(s) &=&\frac{s^3}{\pi}\,\int\frac{dx}{x^{ 3}}\,
\frac{{\mbox{Im}}F_S^{\pi^0}\!(x)/F_S^{\pi^0}\!(0)}{x - s -i0}
\nonumber\\
U_S^{\pi}(s) &=&  \frac{s^3}{\pi}\,\int\frac{dx}{x^{ 3}}\,
\frac{{\mbox{Im}}F_S^{\pi}(x)/F_S^{\pi}(0)}{x - s -i0}
\nonumber\\
U_V^{\pi}(s) &=& \frac{s^3}{\pi}\,\int\frac{dx}{x^{  3}}\,
\frac{{\mbox{Im}}F_V^{\pi}(x)}{x - s -i0}\,.
\end{eqnarray}
For the scattering amplitudes, we start from fixed-$t$ dispersion relations with three subtractions \cite{Stern:1993rg}
\begin{eqnarray}
{A}(s,t) = {P}(t \vert s,u) + \frac{s^3}{\pi} \int \frac{d x}{x^3} \frac{1}{x-s-i0} \, {\mbox{Im}}_s  A (x,t)
+ \frac{u^3}{\pi} \int \frac{d x}{x^3} \frac{1}{x-u-i0} \,  {\mbox{Im}}_u {A}(x,t)
.
\label{disp_amp}
\end{eqnarray}
In the case where $M_\pi \neq M_{\pi^0}$, one has several amplitudes
to consider \cite{DescotesGenon:2012gv}, according to the number of
charged pions involved. Eq. (\ref{disp_amp}) merely displays the
general structure of the corresponding dispersion relations.
The absorptive parts in the $s$ and $u$ channels are related by crossing.

The second ingredient consists of the partial wave expansions of the $\pi\pi$ amplitudes 
[this is one instance where the case of the pion form factors is simpler:
the $K_{\ell 4}$ form factors, which depend on an angular variable, are
also subject to a decomposition into partial-wave projections, see below]

\begin{equation}
A(s,t) \,=\, 16\pi \sum_{l\ge 0} (2\ell +1) P_{l}(\cos\theta) f_{l}(s) 
,
\qquad
f_l (s) \,=\, \frac{1}{32\pi} \int_{-1}^{+1} dz A(s,t) P_l (z) 
.
\end{equation}

The third ingredient is provided by chiral counting for the 
partial waves and the form factors. If $E$ denotes a pion momentum or a pion mass,
the leading behaviour of the various quantities is given by
\begin{eqnarray}
&{\mbox{Re}}F_S^{\pi(\pi^0)}(s) \sim {\cal O}(E^2),\quad  
&{\mbox{Im}}F_S^{\pi(\pi^0)}(s) \sim {\cal O}(E^4),
\nonumber\\
&{\mbox{Re}}F_V^{\pi}(s) \sim {\cal O}(E^0),\quad  
&{\mbox{Im}}F_V^{\pi}(s) \sim {\cal O}(E^2) 
,
\label{counting1}
\end{eqnarray}
and
\begin{eqnarray}
%\!\!\!\!\!\!\!\!
&&
{\mbox{Re}}f_{l}(s)  \sim {\cal O}(E^2),
\, {\mbox{Im}}f_{l}(s)  \sim {\cal O}(E^4),
\, l = 0,1 , 
\nonumber\\
&&
{\mbox{Re}}f_{l}(s)  \sim {\cal O}(E^4),
\, {\mbox{Im}}f_{l}(s)  \sim {\cal O}(E^8),
\, l \ge 2 
.
\label{counting2}
\end{eqnarray}
These properties allow us to decompose the real parts of the $l = 0,1$ partial waves as
\begin{equation} 
{\mbox{Re}} f_l (s) = \varphi_l (s) + \psi_l (s) + {\cal O}(E^6) 
,\quad
\varphi_l (s) \sim {\cal O}(E^2)
,\quad
\psi_l (s) \sim {\cal O}(E^4)
,
\end{equation}
so that
\begin{equation}
\left\vert f_l (s) \right\vert ^2 \,=\, \left[ {\mbox{Re}} f_l (s) \right]^2 
\,+\, {\cal O}(E^8)\,=\, \left[ \varphi_l (s) \right]^2 + 2 \varphi_l (s) \psi_l(s)
\,+\, {\cal O}(E^8)  , \ l=0,1 
.
\end{equation}

Analyticity and unitarity, which together make up the fourth ingredient,
give us information about the cut singularities and their discontinuities.
The absorptive parts of the dispersion relations we started with are given by unitarity. 
These discontinuities are restricted by power counting.
Indeed, in the low-energy region, only two-pion intermediate states
occur up to two loops. Making use of the counting rules in Eqs. (\ref{counting1})
and (\ref{counting2}), we are then led to
\begin{eqnarray}
{\mbox{Im}}F_S^{\pi^0}(s) &=& {\mbox{Re}}\bigg\{
\frac{1}{2}\,\sigma_{0}(s)f_0^{00}(s) F_S^{\pi^0\! *}(s) \theta(s-4M_{\pi^0}^2)
%\nonumber\\
%&&\!\!\!\!\!\!\!\!\!\! 
-
\sigma (s)f_0^{x}(s)F_S^{\pi *}(s) \theta(s-4M_{\pi}^2)
\bigg\} + {\cal O}(E^8) ,
\nonumber\\
{\mbox{Im}}F_S^{\pi}(s) &=& {\mbox{Re}}\bigg\{
\sigma (s)f_0^{\mbox{\tiny{$ +-$}}}(s)F_S^{\pi *}(s)\theta(s-4M_{\pi}^2)
%\nonumber\\
%&&\!\!\!\!\!\!\!\!\!\! 
- 
\frac{1}{2}\,\sigma_{0}(s)f_0^{x}(s)F_S^{\pi^0\! *}(s)\theta(s-4M_{\pi^0}^2)
\bigg\} + {\cal O}(E^8) ,
\nonumber\\
{\mbox{Im}}F_V^{\pi}(s) &=& {\mbox{Re}}\bigg\{
\sigma (s)f_1^{\mbox{\tiny{$ +-$}}}(s)F_V^{\pi *}(s)\theta(s-4M_{\pi}^2) \bigg\}
%\nonumber\\
%&&\!\!\!\!\!\!\!\!\!\! 
+ {\cal O}(E^6) ,
\nonumber\\
\nonumber\\
{\mbox{Im}} A(s,t) &=& 16\pi \left[ {\mbox{Im}} f_0(s) + 3 z {\mbox{Im}} f_{1}(s) \right] + {\cal O}(E^8)
%\underbrace{\Phi_{\ell\ge 2}(s,t) }_{\sim {\cal O}(E^8)}
.
\end{eqnarray}
%\end{widetext}
%where we define the phase-space functions
Here,
\begin{eqnarray}
\sigma_{0}(s)\,=\,\sqrt{1 - \frac{4M_{\pi^0}^2}{s}}\,,\ \sigma(s)\,=\,\sqrt{1 - \frac{4M_{\pi}^2}{s}} 
\end{eqnarray}
denote the  neutral and charged two-pion phase spaces.

We possess now all the tools necessary to proceed towards
the construction of the two-loop representations of the
form factors and scattering amplitudes through an iterative
two-step process that is described by Fig. \ref{fig1} below.
\begin{figure}
%\begin{center}
\begin{picture}(375,130)(0,130)

\SetWidth{1.5}

\Boxc(0,200)(70,20)
\Text(0,200)[]{$A$ at order {$E^{2k}$}}

\ArrowLine(35,200)(90,200)

\Text(62.5,240)[]{projection}
\Text(62.5,230)[]{over partial waves}

\Boxc(125,200)(70,20)
\Text(125,200)[]{$f$ at order {$E^{2k}$}}

\ArrowLine(160,200)(202.5,200)

\Text(180,230)[]{unitarity}

\Boxc(250,200)(95,20)
\Text(250,200)[]{Im $f$ at order {$E^{2k+2}$}}

\ArrowLine(297.5,200)(335,200)

\Text(317.5,230)[]{dispersion relation}

\Boxc(375,200)(80,20)
\Text(375,200)[]{$A$ at order {$E^{2k+2}$}}

\ArrowLine(375,190)(375,150)
\ArrowLine(375,150)(0,150)
\ArrowLine(0,150)(0,190)

\end{picture}
%\end{center}
\caption{Schematic display of the recursive construction of two-loop representations
for the form factors and the $\pi\pi$ scattering amplitudes in the low-energy regime.
$A$ denotes the amplitude of interest, whereas $f$ corresponds to partial waves.}
\label{fig1}
\end{figure}
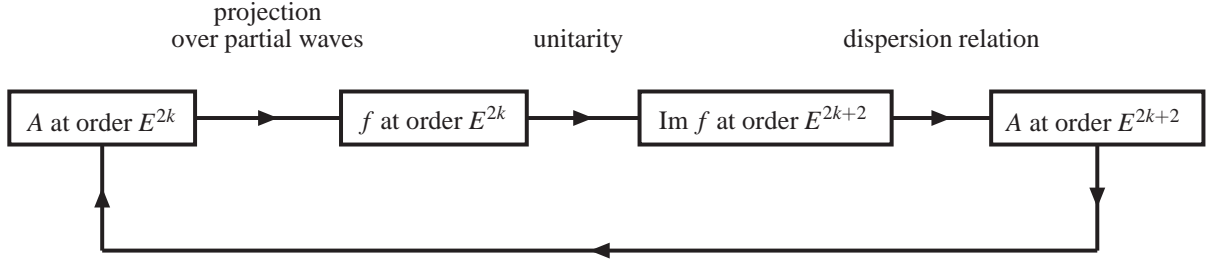
We start with the expressions of the $\pi\pi$ amplitudes at
lowest order [the superscript $00$ stands for $\pi^0\pi^0 \to \pi^0\pi^0$,
$+-$ for $\pi^+\pi^- \to \pi^+\pi^-$, and $x$ for the inelastic
$\pi^+\pi^- \to \pi^0\pi^0$ channel]
\begin{equation}
 A^{00}(s,t) = 16 \pi a_{00} ,
\quad
A^x (s,t) = 16 \pi \left[ a_x + b_x \frac{s - 4 M_\pi^2}{F_\pi^2} \right] ,
\quad
A^{\mbox{\tiny{$ +-$}}} (s,t) =
16 \pi \left[ a_{\mbox{\tiny{$ +-$}}} + b_{\mbox{\tiny{$ +-$}}} \frac{s - 4 M_\pi^2}{F_\pi^2} 
+ c_{\mbox{\tiny{$ +-$}}} \frac{t-u}{F_\pi^2}\right]
,
\end{equation}
from which we obtain the partial-wave projections
\begin{equation}
 \varphi_0^{00}(s) =  a_{00} ,
\quad
\varphi_0^x (s) = a_x + b_x \frac{s - 4 M_\pi^2}{F_\pi^2} ,
\quad
\varphi_0^{\mbox{\tiny{$ +-$}}} (s) = a_{\mbox{\tiny{$ +-$}}} + b_{\mbox{\tiny{$ +-$}}} \frac{s - 4 M_\pi^2}{F_\pi^2}  ,
\quad
\varphi_1^{\mbox{\tiny{$ +-$}}} (s) = c_{\mbox{\tiny{$ +-$}}} \frac{s - 4 M_\pi^2}{F_\pi^2}
.
\end{equation}
This input then gives us the absorptive parts of the one-loop form factors
\begin{eqnarray}
{\mbox{Im}}F_S^{\pi^0}(s) &=&  
\frac{1}{2}\,\sigma_{0}(s)\varphi_0^{00}(s)F_S^{\pi^0}(0)\theta(s-4M_{\pi^0}^2)
-
\sigma (s)\varphi_0^{x}(s)F_S^{\pi}(0)\theta(s-4M_{\pi}^2)
\,+\,{\cal O}(E^6)
\nonumber\\
\nonumber\\
{\mbox{Im}}F_S^{\pi}(s) &=&  
\sigma (s)\varphi_0^{\mbox{\tiny{$ +-$}}}(s)F_S^{\pi}(0)\theta(s-4M_{\pi}^2)
-
\frac{1}{2}\,\sigma_{0}(s)\varphi_0^{x}(s)F_S^{\pi^0}(0)\theta(s-4M_{\pi^0}^2)
\,+\,{\cal O}(E^6)
\nonumber\\
\nonumber\\
{\mbox{Im}}F_V^{\pi}(s) &=&  
\sigma (s)\varphi_1^{\mbox{\tiny{$ +-$}}}(s) \theta(s-4M_{\pi}^2) \,+\,{\cal O}(E^4) 
.
\end{eqnarray}
Injecting them into the dispersion relations, we obtain the one-loop expressions of the form factors
as
\begin{eqnarray}
F_S^{\pi^0}(s) &=& F_S^{\pi^0}\! (0)\! \Bigg[
1 + a_S^{\pi^0} \! s + %b_S^{\pi^0} \! s^2 \! 
16\pi \frac{\varphi_0^{00}(s)}{2}  {\bar J}_0 (s) \! \Bigg]
  \,-\,16\pi {F_S^{\pi}(0)}\, \varphi_0^{x}(s)\,{\bar J} (s)
\nonumber\\
F_S^{\pi}(s) &=& F_S^{\pi}(0) \!\Bigg[
1 +  a_S^{\pi}\,s + %b_S^{\pi}\,s^2 \! + 
16\pi \varphi_0^{\mbox{\tiny{$ +-$}}}(s)   {\bar J} (s)
\Bigg]
 \,-\,16\pi {F_S^{\pi^0}(0)} \,\frac{1}{2}\,\varphi_0^{x}(s)\,
{\bar J}_0 (s)
\nonumber\\
F_V^{\pi}(s) &=& 
1 + a_V^{\pi}\,s %+ b_V^{\pi}\,s^2 
\,+\,16\pi \varphi_1^{\mbox{\tiny{$ +-$}}}(s)   {\bar J} (s)
,
\end{eqnarray}
where
\begin{eqnarray}
{\bar J}_0 (s) &=& \frac{s}{16\pi^2}\,\int_{4M_{\pi^0}^2}^{\infty}\,\frac{dx}{x}\,\frac{1}{x-s-i0}\,\sigma_0 (x)
\,=\, \frac{-1}{16\pi^2}\int_0^1 dx \ln\left[1 - x(1-x)\frac{s}{M_{\pi^0}^2}\right]
\nonumber\\
 {\bar J} (s) &=& \frac{s}{16\pi^2}\,\int_{4M_{\pi}^2}^{\infty}\,\frac{dx}{x}\,\frac{1}{x-s-i0}\,\sigma (x)
 \,=\, \frac{-1}{16\pi^2}\int_0^1 dx \ln\left[1 - x(1-x)\frac{s}{M_\pi^2}\right]
 \label{Jbar}
.
\end{eqnarray}
The scattering lengths , $a_{\mbox{\tiny{$+-$}}}$, $a_x$, $a_{00}$,
and the slope parameters $b_{\mbox{\tiny{$+-$}}}$, $b_x$
are related, at this order, to the scattering lengths $a_0^0$ and $a_0^2$ 
in the isospin limit by \cite{Knecht:1997jw,DescotesGenon:2012gv,Bernard:2013faa}
\begin{eqnarray}
&&
a_{\mbox{\tiny{$+-$}}}\,=\,\frac{2}{3}\,a^0_0\,+\,\frac{1}{3}\,a^2_0\,-\,2 a^2_0\,\frac{\Delta_\pi}{M_{\pi}^2}\ ,
\quad
b_{\mbox{\tiny{$+-$}}}\,=\,c_{\mbox{\tiny{$+-$}}} \,=\, \frac{1}{24} \frac{F_\pi^2}{M_\pi^2} \left( 2 a^2_0 - 5 a^2_0 \right)\,,
\nonumber\\
&&
a_x \,=\, - \frac{2}{3}\,a^0_0\,+\,\frac{2}{3}\,a^2_0\,+\, a^2_0\,\frac{\Delta_\pi}{M_{\pi}^2}\ ,
\quad
b_x\,=\,- \frac{1}{12} \frac{F_\pi^2}{M_\pi^2} \left( 2 a^2_0 - 5 a^2_0 \right)\,,
\nonumber\\
&&
a_{00}\,=\,\frac{2}{3}\,a^0_0\,+\,\frac{4}{3}\,a^2_0\,-\,\frac{2}{3}\left(a^0_0 + 2 a^2_0\right)\frac{\Delta_\pi}{M_{\pi}^2} \,,
\qquad
\Delta_\pi \equiv M_{\pi}^2 - M_{\pi^0}^2
.
\end{eqnarray}
The subtraction constants $a_S^{\pi^0}$, $a_S^{\pi}$, and $a_V^{\pi}$ are related
to the corresponding mean-square radii \cite{DescotesGenon:2012gv} in a calculable way,
in terms of the scattering lengths.
The same procedure can be applied to the $\pi\pi$ scattering amplitudes themselves.
Let us just quote the result for the $\pi^0\pi^0 \to \pi^0\pi^0$ case,
\begin{eqnarray}
A^{00}(s,t,u) &=& 
P^{00}(s,t,u) \,+\, W^{00}_0(s) \,+\, W^{00}_0(t) \,+\, W^{00}_0(u) \,+\, {\cal O}(E^8) .
\end{eqnarray}
Here the function $W^{00}_0(s)$ has a discontinuity starting
at $s=4M_{\pi^0}$ along the {\it positive} $s$ axis 
[the function $W^{00}_0(s)$ itself has only a right-hand cut;
the left-hand cut of the amplitude $A^{00}(s,t,u)$ results
from the two other contributions, involving $W^{00}_0(t)$ and
$W^{00}_0(t)$]. At one-loop order it reads
\begin{eqnarray}
\frac{1}{16\pi}\,{\mbox{Im}}W^{00}_0(s) \,=\,
\frac{1}{2}\,\sigma_0 (s)\,
\left[\varphi_0^{00}(s) \right]^2
\theta (s-4M_{\pi^0}^2)
\,+\,
\sigma (s)\,
\left[\varphi_x (s) \right]^2
\theta (s-4M_{\pi}^2) + {\cal O}(E^6)
,
\end{eqnarray}
so that
\begin{equation}
W^{00}_0(s) = \frac{1}{2} \left[ 16 \pi \varphi_0^{00}(s) \right]^2 {\bar J}_0 (s)
+
\left[ 16 \pi \varphi_0^{x}(s) \right]^2 {\bar J} (s)
.
\end{equation}
Finally,
$P^{00}(s,t,u)$ represents a polynomial of at most second order (at one loop) 
in $s,t,u$, symmetric under any permutation of its variables [due to
the fact that $A^{00}(s,t,u)$ transforms into itself under crossing]
\begin{equation}
P^{00}(s,t,u) = 16\pi a_{00} \,-\,w_{00} 
%\nonumber\\
%&& \!\!\!\!\!\!\!
 +\,
\frac{3\lambda_{00}^{(1)}}{F_{\pi}^4}\left[
s(s-4M_{\pi^0}^2) + t(t-4M_{\pi^0}^2) + u(u-4M_{\pi^0}^2)
\right]
.
\end{equation}
$\lambda_{00}^{(1)}$ denotes an additional subtraction constant,
which can be related to two subtraction constants that describe the
$\pi\pi$ amplitude in the isospin limit \cite{Knecht:1995tr}
\begin{equation}
\lambda_{00}^{(1)} = \frac{1}{3} \left( \lambda_1 + 2 \lambda_2 \right)
,
\end{equation}
and whose values are known \cite{Knecht:1995ai,DescotesGenon:2001tn}.
The quantity
\begin{equation}
w_{00} = {\mbox{Re}}\,
\left[W^{00}_0(4M_{\pi^0}^2)\,+\, W^{00}_0(0)\,+\, W^{00}_0(0)\right]
\end{equation}
is then uniquely fixed by the requirement that $a_{00}$ retains its meaning as scattering length
at next-to-leading order, i.e. ${\mbox{Re}} A^{00}(4 M_{\pi^0}^2,0,0) = 16 \pi a_{00}$.
The structure of the other amplitudes is similar, and we refer the interested reader
to \cite{DescotesGenon:2012gv} for details. 

With the one-loop form factors
and amplitudes at our disposal, we can
now repeat the same procedure: compute the
$S$ and $P$ partial-wave projections from the one-loop
amplitudes, use them to express the discontinuities
of the two-loop form factors and amplitudes,
and eventually obtain the full two-loop form
factors and amplitudes. Their expressions will involve a limited
number of additional subtraction constants, which
can however be related to the parameters that
describe the same quantities in the isospin limit,
the IB breaking corrections being expressed in
terms of the scattering lengths \cite{DescotesGenon:2012gv}.
The remarkable feature of this 
second iteration is that the partial-wave projections
of the one-loop amplitudes
can be obtained analytically, using the known expressions
of the functions ${\bar J}_0 (s)$ and ${\bar J} (s)$ in terms 
of elementary functions. However, it is in general
not possible to perform all the corresponding dispersion integrals
analytically if $M_\pi \neq M_{\pi^0}$, in contrast to
the situation in the isospin limit, where analytical expressions are 
available \cite{Knecht:1995tr}. Thus, the real parts of the two-loop
amplitudes and form factors are partly known only as one-dimensional
integrals, which have to be evaluated numerically. However, the
expressions of the phases at two loops only involve the real
parts at one loop, which are known analytically. We have therefore
reached our goal, in this somewhat simpler setting, of obtaining expressions
of the phases at two-loop precision, parameterized in terms of the
scattering lengths in the isospin limit. We will now briefly
explain how essentially the same procedure can be used in order
to obtain two-loop expressions for the (phases of the) $K_{e4}$ form
factors that depend parametrically on the scattering lengths.

\section{Two-loop representation of $K_{e4}$ form factors with IB}

The construction of two-loop representations for the form factors 
describing the matrix elements for the $K_{\ell 4}$ transitions $K^\pm\to\pi^+\pi^- \ell^\pm \nu_\ell$,
$\ell = e , \mu$, proceeds essentially along the same lines. On the technical
level, additional complications arise, due, on the one hand, to the fact that
there are several form factors, related by crossing, to consider simultaneously,
and, on the ohter hand, that these form factors depend on two energy variables
and one angular variable. In this Section, we will successively go through the list
of ingredients listed in the preceding Section, and describe the changes
that are induced by these two features.

In the Standard Model, the amplitudes corresponding to $K_{\ell 4}$ decays are defined by the 
matrix elements of the type $\langle \pi^a(p_a) \pi^b(p_b) \vert i A_{\mu} (0) \vert K(k) \rangle$
and $\langle \pi^a(p_a) \pi^b(p_b) \vert i V_{\mu} (0) \vert K(k) \rangle$
involving the $\Delta S = \Delta Q = +1$ axial and vector currents between a (charged or
neutral) kaon state and the corresponding two-pion state,  specifically
$(K,a,b)\in\{(K^+,+,-),(K^+,0,0),(K^0,0,-)\}$. In the present study, we will not 
consider the matrix element of the vector current,
related to the axial anomaly, and described by a single form factor $H^{ab}(s,t,u)$.
Since crossing is one of the ingredients of our construction, we also need to consider the matrix elements related to
$\langle \pi^a(p_a) \pi^b(p_b) \vert i A_{\mu} (0) \vert K(k) \rangle$ through this operation, namely
$\langle \pi^a(p_a) {\bar K}(k) \vert iA_{\mu} (0) \vert {\bar \pi}^b(p_b)  \rangle $ and
$\langle {\bar K}(k) \pi^b(p_b) \vert iA_{\mu} (0) \vert {\bar \pi}^a(p_a) \rangle $.
In order to be able to treat these matrix elements 
simultaneously and on a common footing, we consider general matrix elements of the type \cite{Bernard:2013faa} 
\begin{equation}
{\cal A}^{ab}_\mu (p_a, p_b ; p_c) = \langle a(p_a) \, b(p_b) \vert iA_\mu (0) \vert {\bar c}(p_c) \rangle
,
\label{A_ab}
\end{equation}
with $\{ a,b,c \} = \{ \pi^+, \pi^-, K^- \}$, $\{ \pi^0, \pi^0, K^- \}$ or $\{ \pi^0, \pi^-, {\bar K}_0 \}$.
These matrix elements possess the general decompositions into invariant form factors
\begin{equation}
{\cal A}^{ab}_\mu (p_a, p_b ; p_c) =
(p_a + p_b)_\mu F^{ab}(s,t,u) +
(p_a - p_b)_\mu G^{ab}(s,t,u) +
(p_c - p_a - p_b)_\mu R^{ab}(s,t,u) .
\label{decomp_A_mu}
\end{equation}
They depend on the variables
$s = (p_a + p_b)^2 ,\ t = (p_c - p_a)^2 ,\ u = (p_c - p_b)^2$, 
obeying the ``mass-shell'' condition
$ s + t + u \,=\, M_a^2 + M_b^2 + M_c^2 + s_{\ell} \equiv \Sigma_\ell$, 
with $\ s_{\ell} \equiv (p_c - p_a - p_b)^2$
being the square of the dilepton invariant mass.
In the physical region of the $K_{\ell 4}$ decay,
$s_{\ell}$ is strictly positive, $s_\ell \ge m_{\ell}^2$, and in what follows we will always assume this to
be the case. Independent variables will conveniently be chosen as $s$, $s_\ell$, and the angle $\theta_{ab}$ 
made by the line of flight of particle $a$ in the $(a,b)$ rest frame with the
direction of ${\vec p}_a + {\vec p}_b$ in the rest frame of particle ${\bar c}$,
\begin{equation}
\cos\theta_{ab}\,=\, \frac{(M_a^2 - M_b^2)(s_\ell -M_c^2) - s(t-u)}{\lambda^{1\over 2}_{ab}(s) \lambda^{1\over 2}_{\ell c} (s)}
\,=\, \frac{(M_a^2 - M_b^2)(s_\ell -M_c^2) + s(\Sigma_\ell - s - 2t)}{\lambda^{1\over 2}_{ab}(s) \lambda^{1\over 2}_{\ell c} (s)}
.
\label{cos_theta}
\end{equation}
The functions $\lambda_{ab}(s)$
and $\lambda_{\ell c} (s)$ are defined in terms of K\"allen's function $\lambda(x,y,z)=x^2+y^2+z^2-2xy-2xz-2yz$
by $\lambda_{ab}(s)=\lambda(s,M_a^2,M_b^2)$ and $\lambda_{\ell c} (s) = \lambda(s,s_\ell,M_c^2)$,
respectively.

As in the case of the pion form factors discussed in the preceding Section,
the starting point of the construction consists of suitably subtracted
dispersion relations for fixed $t$ {\it and} $s_\ell$. Before writing
down the relevant dispersion relations, let us briefly discuss under which form the other 
ingredients that were listed and used in the case of the pion form factors
enter in the present case.

\indent

\noindent
$\bullet$~Crossing properties

In the previous Section, crossing was only relevant for
the $\pi\pi$ scattering amplitudes. In the present
case, the matrix elements (\ref{A_ab}) are also concerned.
Their crossing properties are expressed through the relations
\begin{eqnarray}
{\cal A}_{\mu}^{ac}(p_a,p_c;p_b) \,=\, \lambda_b \lambda_c {\cal A}_{\mu}^{ab}(p_a,-p_b;-p_c)
\, ,\quad
{\cal A}_{\mu}^{cb}(p_c,p_b;p_a) \,=\, \lambda_a \lambda_c {\cal A}_{\mu}^{ab}(-p_a,p_b;-p_c)
,
\end{eqnarray}
where the matrix elements on the right-hand sides are related through analytic 
continuations
to the original matrix element ${\cal A}_{\mu}^{ab}(p_a,p_b;p_c)$,
assuming that the usual analyticity properties hold.
The coefficients $\lambda_{a,b,c}$ are crossing phases, which 
are chosen such as to reduce to the Condon-Shortley phase convention in the isospin limit,
\begin{equation}
\lambda_{K^\pm}\,=\,\lambda_{\pi^\pm} \,=\, -1,
\ \lambda_{\pi^0} \,=\, \lambda_{K^0} \,=\, \lambda_{{\bar K}^0} \,=\, +1 .
\end{equation}
At the level of the form factors themselves, these crossing relations become
\begin{equation}
%\!\!\!\!\!\!\!\!\!\!
{\bf A}^{ac}(s,t,u)\,=\,\lambda_b \lambda_c\, {\cal C}_{st} {\bf A}^{ab}(t,s,u)\,,
\ {\bf A}^{cb}(s,t,u)\,=\,\lambda_a \lambda_c\, {\cal C}_{us} {\bf A}^{ab}(u,t,s)\,,
\ {\bf A}^{ba}(s,t,u)\,=\,{\cal C}_{tu} {\bf A}^{ab}(s,u,t)
,
\\
\label{crossing}
\end{equation}
with
\begin{eqnarray}
{\bf A}^{X}(s,t,u) \,=\, \left(
\begin{array}{c}
F^{X}(s,t,u) \\
G^{X}(s,t,u) \\
R^{X}(s,t,u)
\end{array}
\right)
,
\label{A^X_three}
\end{eqnarray}
where $X$ stands for any one of the couples of indices $ab$ (and, in the present case, also $ba$),
$ac$, or $cb$,
and
\begin{eqnarray}
{\cal C}_{st} \,=\,  \left(
\begin{array}{ccc}
-\frac{1}{2} & +\frac{3}{2} & 0 \\
+\frac{1}{2} & +\frac{1}{2} & 0 \\
-1           &    +1        & +1\\
\end{array}
\right)
\,,\ {\cal C}_{us} \,=\,  \left(
\begin{array}{ccc}
-\frac{1}{2} & -\frac{3}{2} & 0 \\
-\frac{1}{2} & +\frac{1}{2} & 0 \\
-1           &    -1        & +1\\
\end{array}
\right)
\,,\ {\cal C}_{tu} \,=\, \left(
\begin{array}{ccc}
+1 & 0 & 0 \\
0 & -1 & 0 \\
0           &    0        & +1\\
\end{array}
\right)
\,.
\end{eqnarray}
Each of these crossing matrices squares to the identity matrix. In addition, they satisfy  the relations
\begin{equation}
{\cal C}_{st}{\cal C}_{us} \,=\, {\cal C}_{us}{\cal C}_{tu},
\ {\cal C}_{us}{\cal C}_{st} \,=\, {\cal C}_{st}{\cal C}_{tu},
\ {\cal C}_{st}{\cal C}_{tu} \,=\, {\cal C}_{tu}{\cal C}_{us}
.
\end{equation}
It is useful to notice that under crossing the form factors 
$F^X$ and $G^X$ transform into form factors $F^Y$ and $G^Y$, without
mixing with the form factors $R^Y$. In the following, we will omit the form factors  $R^X$ from the discussion most of the time, writing
\begin{eqnarray}
{\bf A}^{X}(s,t,u) \,=\, \left(
\begin{array}{c}
F^{X}(s,t,u) \\
G^{X}(s,t,u) 
\end{array}
\right)\,,
\label{A^X_two}
\end{eqnarray}
instead of Eq.~(\ref{A^X_three}). When it is the case, it is understood that the crossing matrices 
are reduced to their upper-left $2\times 2$ blocks.
All the previous relations between these matrices remain unaffected by this truncation.

\indent

\noindent
$\bullet$~Partial-wave projections

The form factors appearing in the decomposition (\ref{decomp_A_mu}) are free from kinematical singularities,
but do not have simple decompositions into partial waves. For the latter, it is more
convenient to introduce another set of form factors. To this effect, adapting the method
of Ref.~\cite{Berends:1968zz} to the more general situation at hand, we define
\begin{eqnarray}
{\cal F}^{ab}(s,t,u) &=& F^{ab}(s,t,u) + \left[\frac{M_a^2 - M_b^2}{s} \,+\, \frac{M_c^2 - s - s_\ell}{s}\,
\frac{\lambda^{1\over 2}_{ab}(s)}{\lambda^{1\over 2}_{\ell c} (s)}\, \cos\theta_{ab} \right] G^{ab}(s,t,u)
,
\nonumber\\
{\cal G}^{ab}(s,t,u) &=& G^{ab}(s,t,u)
,
\nonumber\\
{\cal R}^{ab}(s,t,u) &=& R^{ab}(s,t,u) +\,\frac{M_c^2-s-s_\ell}{2 s_\ell}\,F^{ab}(s,t,u)
\nonumber\\
&&
 +\, \frac{1}{2 s s_\ell} \left[
(M_a^2 - M_b^2)(M_c^2 - s - s_\ell) + \lambda^{1\over 2}_{ab}(s) \lambda^{1\over 2}_{\ell c} (s) \cos\theta_{ab}
\right] G^{ab}(s,t,u).
\label{F_G_R_cal}
\end{eqnarray}
Notice that the form factor ${\cal R}^{ab}(s,t,u)$ describes the matrix element of the
divergence of the current $A^\mu (x)$,
\begin{equation}
\langle a(p_a) \, b(p_b) \vert \partial^\mu A_\mu (0) \vert {\bar c}(p_c) \rangle \,=\, -s_\ell {\cal R}^{ab}(s,t,u)
\label{calR:div_A}
.
\end{equation}
These form factors have the following partial-wave decompositions  \cite{Berends:1968zz}
\begin{eqnarray} 
{\cal F}^{ab}(s,t,u) &=& \sum_{l \ge 0} f_{l}^{ab}(s,s_{\ell}) P_{l}(\cos\theta_{ab}) 
,
\nonumber\\
{\cal G}^{ab}(s,t,u) &=& \sum_{l \ge 1} g_{l}^{ab}(s,s_{\ell}) P_{l}^{\prime}(\cos\theta_{ab})
,
\nonumber\\
{\cal R}^{ab}(s,t,u) &=& \sum_{l \ge 0} r_{l}^{ab}(s,s_{\ell}) P_{l}(\cos\theta_{ab})
.
\label{FandG_PW_decomp}
\end{eqnarray}
%where $P_{l}^1(\cos\theta)=\sin\theta P_{l}^{\prime}(\cos\theta)$.
Since $\{F ; G ; R\}^{ab} (s , t , u) = \{F ; - G ; R\}^{ba} (s , u , t)$ and $\cos\theta_{ab} = - \cos\theta_{ba}$,
one has the symmetry properties
\begin{equation}
f_{l}^{ba} (s,s_{\ell}) = (-1)^l f_{l}^{ab} (s,s_{\ell})\,, 
\ g_{l}^{ba} (s,s_{\ell}) = (-1)^{l} g_{l}^{ab} (s,s_{\ell}) \,, 
\ r_{l}^{ba} (s,s_{\ell}) = (-1)^l r_{l}^{ab} (s,s_{\ell}) 
.
\label{sym_PV}
\end{equation}
Let us also note that the form factors ${\cal F}^X$ and ${\cal G}^X$ transform among themselves under crossing.
On the other hand, and in contrast with the form factors $R^X$,
the form factors ${\cal R}^X$ transform into themselves, without mixing with ${\cal F}^X$ and ${\cal G}^X$,
\begin{eqnarray}
{\cal R}^{ac}(s,t,u) \,=\, \lambda_b\lambda_c {\cal R}^{ab}(t,s,u)\,,
\ {\cal R}^{cb}(s,t,u) \,=\, \lambda_a\lambda_c {\cal R}^{ab}(u,t,s)
.
\end{eqnarray}
This result follows from Eq.~(\ref{calR:div_A}): the form factors ${\cal R}^X$ cannot mix under crossing with the other form factors,
which correspond to the transverse components of axial current.

\indent

\noindent
$\bullet$~Chiral counting

The chiral counting is given by $M_P \sim {\cal O}(E)$, $s,t,u,s_\ell \sim {\cal O}(E^2)$, where $M_P=M_\pi , M_{\pi^0} , M_K$,
and $s_\ell$ is treated on the same footing as one of the masses squared, which is
compatible with its allowed range inside the $K_{\ell 4}$ phase space. On the level of the
partial waves, this gives \cite{Colangelo:1994qy} [the counting of the $\pi\pi$ partial waves remains of course unchanged]
\begin{eqnarray}
&{\mbox{Re}}f_0^{ab}(s, s_\ell),\ {\mbox{Re}}f_1^{ab}(s,s_\ell),\ {\mbox{Re}\,}g_1^{ab}(s,s_\ell) \sim {\cal O}(E^0)),\quad
&{\mbox{Im}}f_0^{ab}(s,s_\ell),\ {\mbox{Im}}f_1^{ab}(s,s_\ell),\ {\mbox{Im}\,}g_1^{ab}(s,s_\ell)  \sim {\cal O}(E^2)
\nonumber
\\
&\!\!\!\!\!\!\!\!\!\!\!\!
{\mbox{Re}}f_l^{ab}(s,s_\ell),\ {\mbox{Re}}\,g_l^{ab}(s,s_\ell) \sim {\cal O}(E^2) ,\ l\ge 2 ,\quad
&{\mbox{Im}}f_l^{ab}(s,s_\ell),\ {\mbox{Im}}\,g_l^{ab}(s,s_\ell)\sim {\cal O}(E^6) ,\ l\ge 2
.
\end{eqnarray}
The $S$ and $P$ waves as therefore dominant at low energies, which makes them the central subject of study for $K_{\ell 4}$ decays.
In terms of the form factors $F^{ab}(s,t,u)$ and $G^{ab}(s,t,u)$, the chiral counting of the partial
waves translates into the decompositions
\begin{eqnarray}
F^{ab}(s,t,u) &=& F_S^{ab}(s,s_\ell)\,+\,F_P^{ab} (s,s_\ell) \cos\theta_{ab} \,+\,F^{ab}_>(s,\cos\theta_{ab},s_\ell)
,
\nonumber
\\
G^{ab}(s,t,u) &=& G_P^{ab}(s,s_\ell)  \,+\,G^{ab}_>(s,\cos\theta_{ab},s_\ell)
.
\label{decomp_F_G}
\end{eqnarray}
The contributions of the partial waves with $l \ge 2$ are collected in $F^{ab}_>(s,\cos\theta_{ab},s_\ell)$ and in 
$G^{ab}_>(s,\cos\theta_{ab},s_\ell)$, with the counting
${\mbox{Re}} F^{ab}_>(s,\cos\theta_{ab},s_\ell)$, ${\mbox{Re}} G^{ab}_>(s,\cos\theta_{ab},s_\ell) \sim {\cal O}(E^2)$ and
${\mbox{Im}} F^{ab}_>(s,\cos\theta_{ab},s_\ell)$, ${\mbox{Im}} G^{ab}_>(s,\cos\theta_{ab},s_\ell)$ $\sim {\cal O}(E^6)$, 
while the contributions from $S$ and $P$ waves are collected in 
\begin{eqnarray}
F_S^{ab}(s,s_\ell) &=& f_0^{ab}(s,s_\ell) \,-\,\frac{M_a^2 - M_b^2}{s}\,g_1^{ab}(s,s_\ell)
,
\nonumber\\
F_P^{ab}(s,s_\ell) &=& f_1^{ab}(s,s_\ell) \,-\, \frac{M_c^2 - s - s_{\ell}}{s}\,
\frac{\lambda^{\frac{1}{2}}_{ab}(s)}{\lambda^{\frac{1}{2}}_{\ell c}(s)}\,g_1^{ab}(s,s_\ell)\,,
\nonumber
\\
G_P^{ab}(s,s_\ell) &=& g_1^{ab}(s,s_\ell)
.
\label{def_F_S_and_F_P}
\end{eqnarray}

\indent

\noindent
$\bullet$~Analyticity and unitarity

We now assume that the form factors $F^{ab}(s,t,u)$ and $G^{ab}(s,t,u)$ have the usual analyticity properties
with respect to the variable $s$, for fixed values of $t$ and of $u$, %[and for $0 \le s_\ell \le (M_K - 2 M_\pi)^2$], 
with a cut on the positive $s$-axis, whose discontinuity is fixed by unitarity, and a cut on the negative
$s$-axis generated by unitarity in the crossed channel. The form factors
are regular and real in the interval between $s=0$ and the positive value of $s$ corresponding
to the lowest-lying intermediate state.
We can thus write the following dispersion relations %[momentarily omitting the dependence with respect to $u$ and/or $s_\ell$]
\begin{equation}
{\bf A}^{ab}(s,t) = {\bf P}^{ab}(t \vert s,u) + \frac{s^2}{\pi} \int \frac{d x}{x^2} \frac{1}{x-s-i0} \, {\mbox{Im}} {\bf A}^{ab}(x,t)
+ \frac{u^2}{\pi} \int \frac{d x}{x^2} \frac{1}{x-u-i0} \, \lambda_a \lambda_c {\cal C}_{us} {\mbox{Im}} {\bf A}^{cb}(x,t)
.
\label{Disp_Rel}
\end{equation}
Each integral runs slightly above or below the corresponding cut in the complex $s$-plane, 
from the relevant threshold, $s_{ab}$ or $u_{ab}$, to infinity.
${\bf P}^{ab}(t \vert s,u)$ denotes a pair of subtraction functions that are polynomials
of the first degree in $s$ and $u$, with coefficients given by arbitrary functions of $t$.
Using the decompositions Eqs.~(\ref{cos_theta}) and (\ref{decomp_F_G}), we may write
\begin{equation}
 {\mbox{Im}} {\bf A}^{ab}(s,t) \,=\, \left(
\begin{array}{l}
{\mbox{Im}} F^{ab}_S(s) + 
{\displaystyle{\frac{s(\Sigma_\ell -s - 2t)-(M_a^2 - M_b^2)(M_c^2 - s_\ell)}{\lambda^{1\over 2}_{ab}(s) \lambda^{1\over 2}_{\ell c} (s)}}}
\, {\mbox{Im}} F^{ab}_P(s)\\
{\mbox{Im}} g_1^{ab}(s)
\end{array}
\right)
+ {\mbox{Im}} {\bf A}^{ab}(s,t)_{l\ge 2}
,
\end{equation}
where $F^{ab}_S(s)$ and $F^{ab}_P(s)$ are given in terms of the lowest partial waves by Eq.~(\ref{def_F_S_and_F_P}).
Furthermore, ${\mbox{Im}} {\bf A}^{ab}(s,t)_{l\ge 2}$ collects the contributions of the higher ($l\ge 2$)
partial-wave projections in (\ref{FandG_PW_decomp}), so that at low energies, ${\mbox{Im}} {\bf A}^{ab}(s,t)_{l\ge 2} = {\cal O}(E^6)$. 
The last property is relevant as long as $s$ and $u$ remain below a typical hadronic scale $\Lambda_H \sim 1$ GeV,
but one should remember that the integrals in Eq. (\ref{Disp_Rel}) involving ${\mbox{Im}} {\bf A}^{ab}(x,t)_{l\ge 2}$ run up to infinity. 
However, in the range of $x$ above $\Lambda_H$, ${\mbox{Im}} {\bf A}^{ab}(x,t)_{l\ge 2} = {\cal O}(E^0)$,
so that (see the similar discussion in Ref.~\cite{Stern:1993rg})
\begin{equation}
\frac{s^2}{\pi} \int \frac{d x}{x^2} \frac{1}{x-s-i0} \, {\mbox{Im}} {\bf A}^{ab}(x,t)_{l\ge 2}
\,=\, \left(\frac{s}{\Lambda_H}\right)^2 {\bf H}^{ab} \,+\, {\cal O}(E^6)
,
\label{HPV_exp}
\end{equation}
where ${\bf H}^{ab}$ denotes a set of constants, whose precise definitions need not concern us here.
We thus obtain the expression
\begin{eqnarray}
 {\bf A}^{ab}(s,t,u) &=&  {\bf P}^{ab}(t \vert s,u) +
\left[ {\bf{\mbox{$\Phi$}}}_{\!{\mbox{\tiny$+$}}}^{ab} (s) - (t-u) {\bf{\mbox{$\Phi$}}}_{\!{\mbox{\tiny$-$}}}^{ab} (s) \right]
 + \lambda_a \lambda_c {\cal C}_{us} \left[ {\bf{\mbox{$\Phi$}}}_{\!{\mbox{\tiny$+$}}}^{cb} (u) - (t-s) {\bf{\mbox{$\Phi$}}}_{\!{\mbox{\tiny$-$}}}^{cb} (u) \right]   
\nonumber\\
&& + \,
\lambda_b \lambda_c {\cal C}_{st} \left[ {\bf{\mbox{$\Phi$}}}_{\!{\mbox{\tiny$+$}}}^{ac} (t) - (s-u) {\bf{\mbox{$\Phi$}}}_{\!{\mbox{\tiny$-$}}}^{ac} (t) \right]
\,+\, {\cal O}(E^6)
.
\label{Disp_Rep}
\end{eqnarray}
In this expression, the pair of functions ${\bf P}^{ab}(t \vert s,u)$  differs 
from the one introduced initially in Eq.~(\ref{Disp_Rel}) in two respects. First, it contains
a contribution that compensates the fourth term on the right-hand side of Eq.~(\ref{Disp_Rep}),
which has been introduced to make the crossing properties manifest. 
Second, the terms of Eq.~(\ref{HPV_exp}) generated by the higher partial waves have also
been absorbed into these polynomials. Therefore,  ${\bf P}^{ab}(t \vert s,u)$ in Eq.~(\ref{Disp_Rep})
still represents a pair of arbitrary polynomials of at most second order in $s$ and $u$,  whose
coefficients are functions of $t$. As for the functions ${\bf{\mbox{$\Phi$}}}_{\!{\mbox{\tiny$\pm $}}}^{ab} (s)$,
they are defined by the fact that they have a cut singularity along the positive real axis,
with discontinuities along this cut expressed
in terms of the lowest partial waves as 
\begin{eqnarray}
 {\mbox{Im}} {\bf{\mbox{$\Phi$}}}_{\!{\mbox{\tiny$+$}}}^{ab} (s) &=& 
\left(
\begin{array}{l}
{\mbox{Im}} f^{ab}_0(s) -  
{\displaystyle{\frac{(M_a^2 - M_b^2)}{\lambda_{\ell c} (s)}}}
\left[
(s - M_c^2 - 3 s_\ell) {\mbox{Im}} g^{ab}_1(s) + 
(M_c^2 - s_\ell) {\displaystyle{\frac{\lambda^{1\over 2}_{\ell c} (s)}{\lambda^{1\over 2}_{ab}(s)}}}{\mbox{Im}} f^{ab}_1(s)
\right]
\\
{\mbox{Im}} g^{ab}_1(s) 
\end{array}
%\!\!
\right)\,,
%\theta (s - s_{ab})
\nonumber\\
{\mbox{Im}} {\bf{\mbox{$\Phi$}}}_{\!{\mbox{\tiny$-$}}}^{ab} (s) &=&
\frac{s}{\lambda^{\frac{1}{2}}_{ab}(s) \lambda^{\frac{1}{2}}_{\ell c}(s)} 
\left(
\begin{array}{l}
{\mbox{Im}} f_1^{ab}(s) \,-\, 
{\displaystyle{\frac{M_c^2 - s - s_\ell}{s}\,\frac{\lambda^{1\over 2}_{ab}(s)}{\lambda^{1\over 2}_{\ell c} (s)}}} {\mbox{Im}} g_1^{ab}(s) \\
0
\end{array}
\right)
%\theta (s - s_{ab})
,
\end{eqnarray}
supplemented by ${\bf{\mbox{$\Phi$}}}_{\!{\mbox{\tiny$\pm $}}}^{ab} (0) = 0$ and by the asymptotic conditions  
\begin{equation}
 \lim_{\vert s \vert \to \infty} \, s^{-3 + \frac{1}{2}(1\mp 1) } \, {\bf{\mbox{$\Phi$}}}_{\!{\mbox{\tiny$\pm$}}}^{ab} (s) = 0
.
\end{equation}
These conditions define ${\bf{\mbox{$\Phi$}}}_{\!{\mbox{\tiny$+$}}}^{ab} (s)$ (${\bf{\mbox{$\Phi$}}}_{\!{\mbox{\tiny $-$}}}^{ab} (s) $)
only up to a polynomial ambiguity, which is of second (first) order in $s$. The contributions of these polynomials
to ${\bf A}^{ab}(s,t,u)$ can then be absorbed by the arbitrary subtraction functions ${\bf P}^{ab}(t|s,u)$
already at hand.
Let us stress once more that the functions ${\bf{\mbox{$\Phi$}}}_{\!{\mbox{\tiny$\pm $}}}^{ab} (s)$
only possess right-hand cuts, with discontinuities specified in terms of those
of the partial waves, whereas the partial-wave projections themselves
in general have a more complicated analytical structure. Enforcing the crossing relations,
one finds that the arbitrary subtraction functions ${\bf P}^{ab}(t|s,u)$ boil down to a pair  of polynomials  
${\bf P}^{ab}(s,t,u)$ of at most second order in 
all three variables $s$, $t$, and $u$, with arbitrary {\it constant} coefficients. These coefficients may 
depend on the masses and on $s_\ell$, in a way that is compatible with the chiral counting. 
The polynomials in the different channels are then related by 
\begin{equation}
{\bf P}^{ac}(s,t,u)\,=\, \lambda_b \lambda_c {\cal C}_{st} {\bf P}^{ab}(t,s,u)\,,
\ {\bf P}^{cb}(s,t,u)\,=\, \lambda_a \lambda_c {\cal C}_{us} {\bf P}^{ab}(u,t,s)\,,
\ {\bf P}^{ba}(s,t,u)\,=\,{\cal C}_{tu} {\bf P}^{ab}(s,u,t)
.
\label{P_crossing}
\end{equation}
Finally, unitarity provides us with the discontinuities of the functions 
${\bf{\mbox{$\Phi $}}}_{\!{\mbox{\tiny$+$}}}^{ab} (s)$ and ${\bf{\mbox{$\Phi $}}}_{\!{\mbox{\tiny $-$}}}^{ab} (s) $:
\begin{eqnarray}
{\mbox{Im}}\,f^{ab}_l (s,s_\ell) 
&=&
\sum_{\{ a^\prime , b^\prime \}} \frac{1}{{\cal S}_{a^\prime b^\prime }}\,\frac{\lambda^{\frac{1}{2}}_{a^\prime b^\prime }(s)}{s}\,
{\mbox{Re}}\,\left\{t_l^{a^\prime b^\prime ; a b}(s) 
\left[ f^{a^\prime b^\prime }_l (s,s_\ell )\right]^{\star} 
\right\} \theta (s - s_{a^\prime b^\prime })
 + {\cal O}(E^8)
 ,
\nonumber\\
{\mbox{Im}}\,g^{ab}_l (s,s_\ell) 
&=&
\sum_{\{ a^\prime , b^\prime \}} \frac{1}{{\cal S}_{a^\prime b^\prime }}\,\frac{\lambda^{\frac{1}{2}}_{a^\prime b^\prime }(s)}{s}\,
\frac{\lambda^{\frac{1}{2}}_{a^\prime b^\prime }(s)}{\lambda^{\frac{1}{2}}_{ab}(s)}\,
{\mbox{Re}}\,\left\{t_l^{a^\prime b^\prime ; a b}(s) 
\left[ g^{a^\prime b^\prime }_l (s,s_\ell)\right]^{\star}
\right\} \theta(s - s_{a^\prime b^\prime })
 + {\cal O}(E^8)
,
\label{Imf_Img}
\end{eqnarray}
where $l=0,1$, and $t_l^{a^\prime b^\prime ; a b}(s)$ denotes the $l$-th partial wave of the
$a^\prime b^\prime \to a b$ scattering amplitude. $s_{a^\prime b^\prime }$ stands for 
the lowest invariant mass squared of the corresponding intermediate state, 
$s_{a^\prime b^\prime } = (M_{a^\prime} + M_{b^\prime})^2$ in terms of the masses
$M_{a^\prime}, M_{b^\prime}$ of the particles in the intermediate state. The symmetry factor reads
${\cal S}_{a^\prime b^\prime } = 1$ in all cases of interest, except for 
$\{ a^\prime , b^\prime \}= \{\pi^0 , \pi^0 \}$ or $\{\eta , \eta\}$, 
where ${\cal S}_{a^\prime b^\prime } = 2$. 

\indent
We have now all the elements in our hands to go through the 
procedure depicted in Fig. \ref{fig1} and obtain first the
one-loop expressions of the form factors, and, from there,
through a second iteration, the two-loop expressions. 
As before, the phases of the various form factors can
be obtained analytically, and the IB contributions can be
expressed in terms of the scattering lengths $a_0^0$ and $a_0^2$.
We will not provide further details here, though. They can
be found in Ref. \cite{Bernard:2013faa}, to which we refer the
interested reader.

\section{Extracting the scattering lengths from data}

\begin{figure}
  \includegraphics[height=.3\textheight]{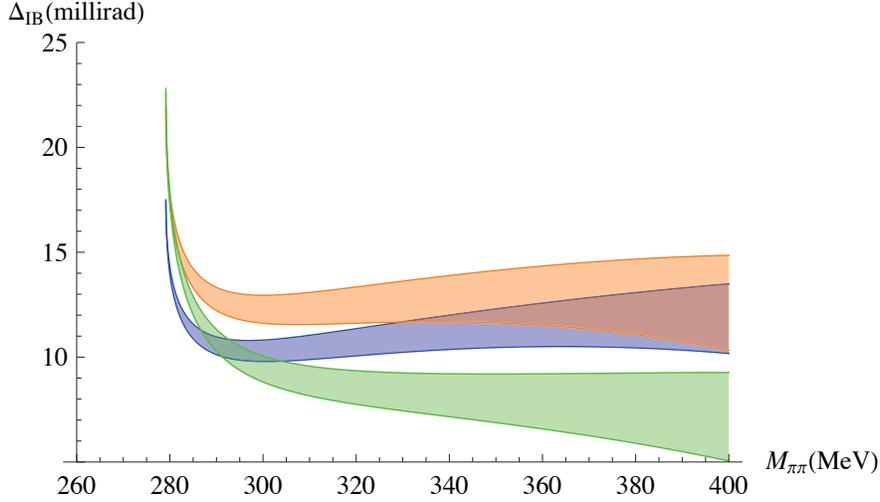}
  \caption{Isospin breaking in the phase of the two-loop form factors, 
$\Delta_{\rm IB}(s,s_\ell) \equiv \delta_{\rm IB} (s ;  a_0^2  , a_0^2  )$
as a function of the dipion invariant mass $M_{\pi\pi}=\sqrt{s}$,
for $s_\ell=0$, for several values of the scattering lengths.
The middle (light-blue) band corresponds to $(a_0^0,a_0^2) = (0.182,-0.052)$,
whereas the other two cases shown correspond to $(a_0^0,a_0^2) = (0.205, -0.055)$ (upper orange band)
and to $(a_0^0,a_0^2) = (0.24, -0.035)$ (lower green band). The widths of these bands result from the
uncertainty on the various inputs needed at two loops. For details, see Ref. \cite{Bernard:2013faa}.}
\label{fig_IB_in_phase_NLO_comb}
\end{figure}

In this Section, we describe how the previous results allow one to analyse the available
phase shifts from {\bf $K_{e4}^\pm$} decays, as provided by the old Geneva-Saclay
experiment \cite{Rosselet:1976pu}, the BNL-E865 experiment ~\cite{Pislak:2001bf}, 
and finally the quite recent NA48/2 experiment~\cite{Batley:2007zz,Batley:2010zza} at the CERN SPS.
Actually, the high accuracy of the latter analysis dominates completely the discussion, 
and we will only consider the data coming  from NA48/2 in the following.
We may restrict the discussion to the two form factors $F (s,t,u)$ and $G (s,t,u)$
[In order to simplify the notation, we suppress the $+-$ superscript, since no confusion can arise]
that occur in the description of the matrix element for the transition
$K^\pm \to \pi^+ \pi^- e^\pm \stackrel{_{(-)}}{\nu_e}$.
The generic low-energy structure of the form factors can be written as in Eq.~(\ref{decomp_F_G}),
\begin{eqnarray}
F(s,t,u) &=& {\widehat F}_S(s,s_\ell) e^{i\delta_S(s, s_\ell)} + {\widehat F}_P(s,s_\ell) e^{i\delta_P(s,s_\ell)} \cos\theta
+ {\mbox{Re}} F_>(s,\cos\theta ,s_\ell) + {\cal O}(E^6)
,
\nonumber\\
G(s,t,u) &=& {\widehat G}_P(s,s_\ell) e^{i\delta_P(s,s_\ell)}
+ {\mbox{Re}} G_>(s,\cos\theta ,s_\ell) + {\cal O}(E^6)
,
\label{F_S_hat_and_G_P_hat}
\end{eqnarray} 
where we have introduced the real functions ${\widehat F}_S(s,s_\ell)$ ($\equiv {\widehat f}_0(s,s_\ell)$ for $M_a = M_b = M_\pi$), 
${\widehat F}_P(s,s_\ell)$, and ${\widehat G}_P(s,s_\ell)$ ($\equiv {\widehat g}_1(s,s_\ell))$, 
which correspond to the quantities appearing in Eq.~(\ref{decomp_F_G}), but with
their phases removed, ${\widehat F}_S(s,s_\ell) =  e^{-i\delta_S(s,s_\ell)} F_S(s+i0,s_\ell)$, etc.
Notice that we have assumed these phases to depend on $s_\ell$, and that 
we have assigned the same phase to $F_P(s,s_\ell)$ and $G_P(s,s_\ell)$.  
The quantity $\left[ \delta_S ( s) - \delta_P (s) \right]_{\rm exp}$ appearing in Eqs. (\ref{data1}),
(\ref{data2}), and (\ref{data_full}) corresponds to the difference $\delta_S(s, s_\ell) - \delta_P(s, s_\ell)$.
In terms of the chiral expansions
\begin{equation}
{\mbox{Re}}\, F_{S}(s,s_\ell) = F_{S[0]} + F_{S[2]}(s,s_\ell) + {\cal O}(E^4) ,
\qquad {\mbox{Re}}\, G_{P}(s,s_\ell) = G_{P[0]} + G_{P[2]}(s,s_\ell) + {\cal O}(E^4)
,
\end{equation}
where $F_{S[0]} , G_{S[0]} \sim {\cal O} (E^0)$ and $F_{S[2]} (s,s_\ell) , G_{S[2]} (s,s_\ell) \sim {\cal O} (E^2)$,
and using the unitarity condition Eq.~(\ref{Imf_Img}) for the imaginary parts, we obtain the expressions
\begin{equation}
\delta_S(s,s_\ell) = 
\sum_{\{ a^\prime , b^\prime \}} \frac{1}{{\cal S}_{a^\prime b^\prime}}\,\frac{\lambda^{\frac{1}{2}}_{a^\prime b^\prime}(s)}{s}\,
\left[
\varphi_0^{a^\prime b^\prime ; {\mbox{\tiny{$+-$}}}} (s) \,
\frac{F_{S[0]}^{a^\prime b^\prime} +  F_{S[2]}^{a^\prime b^\prime} (s,s_\ell)}{F_{S[0]} + F_{S[2]} (s,s_\ell)}
\,
+
\,
\psi_0^{a^\prime b^\prime ; {\mbox{\tiny{$+-$}}}}(s) \,
\frac{F_{S[0]}^{a^\prime b^\prime}}{F_{S[0]}}
\right]
\theta (s - s_{a^\prime b^\prime} ) 
+
{\cal O}(E^6)
,
\label{delta_S}
\end{equation}
and
\begin{equation}
\delta_P(s,s_\ell) = 
\sum_{\{ a^\prime , b^\prime \}} 
\frac{\lambda^{\frac{1}{2}}_{a^\prime b^\prime}(s)}{s}\,
\frac{\lambda^{\frac{1}{2}}_{a^\prime b^\prime}(s)}{\lambda^{\frac{1}{2}}_{ab}(s)}
\left[
\varphi_1^{a^\prime b^\prime ; {\mbox{\tiny{$+-$}}}} (s) \,
\frac{G_{P[0]}^{a^\prime b^\prime} +  G_{P[2]}^{a^\prime b^\prime} (s,s_\ell)}{G_{P[0]} + G_{P[2]} (s,s_\ell)}
\,
+
\,
\psi_1^{a^\prime b^\prime ; {\mbox{\tiny{$+-$}}}}(s) \,
\frac{G_{P[0]}^{a^\prime b^\prime}}{G_{P[0]}}
\right]
\theta (s - s_{a^\prime b^\prime} ) 
+
{\cal O}(E^6)
.
\label{delta_P}
\end{equation}
We see that the phases $\delta_S(s,s_\ell)$ and $\delta_P(s,s_\ell)$ depend on $s_\ell$
through the order ${\cal O}(E^2)$ corrections to the form factors, as soon as a second
intermediate state $a^\prime b^\prime \neq +-$ is involved.
In the case of the $P$-wave phase shift,
there can be no contribution from states with two identical particles due to Bose symmetry,
explaining the absence of 
the factor $1/{{\cal S}_{a^\prime b^\prime}}$ in $\delta_P(s,s_\ell)$.
Hence, for $\delta_P$ in the specific case $ab = +-$ and for $s\le M_K^2$, the sum boils down to the single $\pi^+ \pi^-$ 
intermediate state, the contribution from form factors drops out altogether, and there is no $s_\ell$ dependence left. In other words, 
while Watson's theorem does not apply to the case of the $\delta_S(s , s_\ell)$ phase
shift due to the occurrence of two distinct possible intermediate states 
[$\pi^0 \pi^0$ and $\pi^+ \pi^-$ for $s\le M_K^2$], it
is still operative in the $l=1$ channel. This explains both why the phases of $F_P(s,s_\ell)$
and of $G_P(s , s_\ell)$ are identical, and why this common phase $\delta_P(s)$ actually does not depend on $s_\ell$.
In the isospin limit, the dependence on $s_\ell$ also drops out from $\delta_S(s,s_\ell)$, and Watson's theorem is recovered, 
i.e. the phases tend towards
\begin{equation}
\delta_S(s,s_\ell) \to \, \delta_0(s),
\ \delta_P(s) \to \, \delta_1(s)
\end{equation}
where $\delta_0(s)$ and $\delta_1(s)$ denote the $\pi\pi$ phases in the $l=0$, $I=0$ and $l=1$, $I=1$ channels, respectively.
It appears that the available statistics has not allowed the NA48/2 experiment to identify a dependence of the phases on 
$s_\ell$ \cite{Batley:2007zz,Batley:2010zza}.  Our formalism allows us to 
check that, from the theoretical side, the dependence on $s_\ell$ is indeed sufficiently small, as compared to other sources of error. 
The quantity $\delta^{S-P}_{\rm Roy} (s ; a_0^0 , a_0^2)$ occuring in Eqs. (\ref{data1}),
(\ref{data2}), and (\ref{data_full}) is given by the difference $\delta_0^{\rm Roy}(s) - \delta_1^{\rm Roy}(s)$
of the solutions of the Roy equations. The correction factor obtained from Eqs. (\ref{delta_S}) and (\ref{delta_P}),
\begin{equation}
\delta_{\rm IB} (s ;  a_0^2  , a_0^2  ) = \left[ \delta_S(s,s_\ell=0) - \delta_0(s) \right] -
\left[ \delta_P(s)  - \delta_1(s) \right]
,
\end{equation}
is shown on Fig. \ref{fig_IB_in_phase_NLO_comb} for several illustrative values of $a_0^0$
and $a_0^2$ allowed by the analysis of Roy equations. For the details, especially as far as the numerical
input used for the various parameters is concerned, we refer the reader to the extensive discussion
in Ref. \cite{Bernard:2013faa}. Let us just notice that, despite the uncertainties 
attached to the other parameters that are involved, the correction for larger values of $s$ can 
depend significantly on the values of the scattering lengths, and can, in particular, 
be different from the one-loop estimate performed in Ref.\cite{Colangelo:2008sm}, 
even if order of magnitude and sign are the same.

\begin{figure}[t]
  \includegraphics[height=.3\textheight]{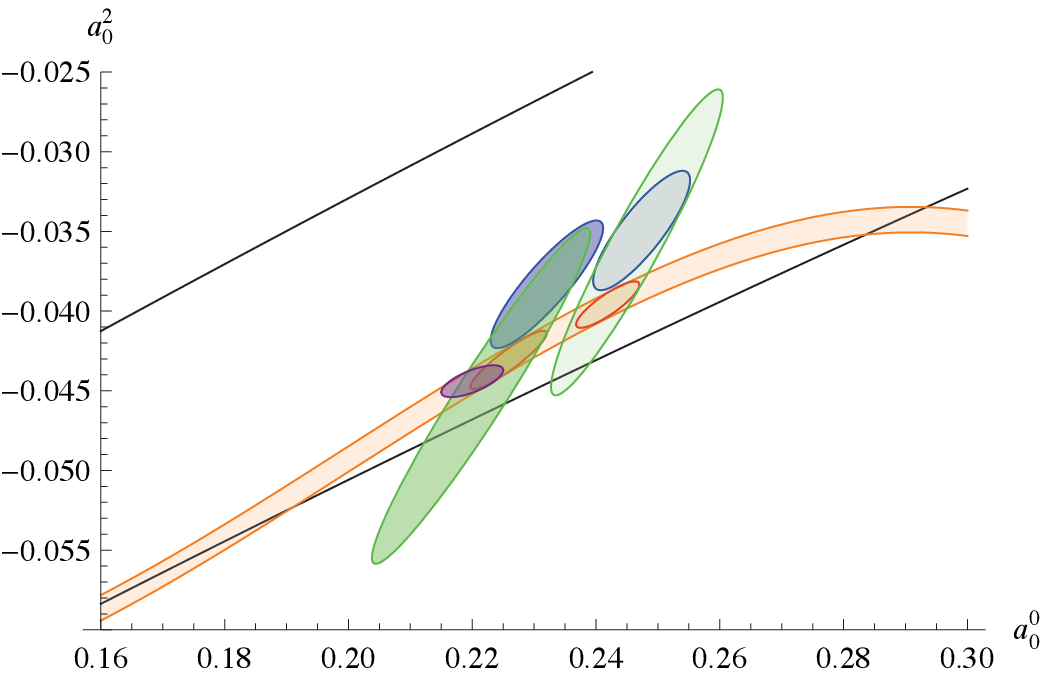}
  \caption{Results of the fits to the NA48/2 data in the $(a_0^0,a_0^2)$ plane. 
The two black solid lines indicate the universal band where the two $S$-wave scattering lengths 
comply with dispersive constraints (Roy equations) and high-energy data on $\pi\pi$ scattering.
The orange band is the constraint  coming from the scalar radius of the pion, cf. Ref.~\cite{Colangelo:2000jc}.
The small dark (purple) ellipse represents 
the prediction based on $N_f=2$ chiral perturbation theory described in Ref.~\cite{Colangelo:2000jc}.
The three other ellipses on the left represent, in order of increasing sizes,
the 1-$\sigma$ ellipses corresponding to the scalar (orange ellipse),  the extended (blue ellipse) and $S$-$P$ (green ellipse) fits,
respectively, when isospin-breaking corrections are included.
The light-shaded ellipses on the right represent the same outputs, but obtained without including 
isopin-breaking corrections.}
\label{fig:fitres}
\end{figure}

The results of our analyses of the NA48/2 data are shown in Fig.~\ref{fig:fitres}, and summarised in Tab.~\ref{tab:fitresults}. 
Note that the NA48/2 data alone lead to a strong correlation between the values of $a_0^0$ and $a_0^2$.
In order to circumvent this problem, we have considered two possible fitting procedures.
The first fit, called extended fit, supplements the NA48/2 data with low-energy data on the $\pi\pi$ scattering
phases in the isospin 2 channel, as described in \cite{DescotesGenon:2001tn}. The second fit,
called the scalar fit, adds a theoretical constraint on the scalar radius of the pion \cite{Colangelo:2000jc}.
We have performed the analysis both in presence and in absence of the isospin-breaking correction terms, 
and we obtain   
\begin{equation}
a_0^0=0.222 \pm 0.013 \,, \qquad a_0^2=-0.043\pm 0.009
.
\end{equation} 
Our result is in good agreement with the one in Eq. (\ref{NA48-2_fit_Model_B}) obtained by
the NA48/2 collaboration for the fit corresponding to the so-called Model B 
in Ref.~\cite{Batley:2010zza},
but with slightly larger errors 
once isospin-breaking corrections are included. This is not surprising since our isospin-breaking 
correction varies with $a_0^0$ and $a_0^2$.
In addition, we notice that the outcome of our fit provides values of $\lambda_1$ and $\lambda_2$ which are compatible with our inputs, 
$\lambda_1=(-4.18\pm 0.63) \cdot 10^{-3}$, $\lambda_2=(8.96\pm 0.12)\cdot 10^{-3}$ -- in agreement with the fact that 
the determination of these two subthreshold parameters has remained very 
stable over time~\cite{Knecht:1995ai,Colangelo:2001df,DescotesGenon:2001tn}.  
We see that in absence of isospin breaking, larger values of 
$a_0^0$ are preferred. %which brings into better agreement the extended fit with the scalar fit.
%Once isospin-breaking corrections are included, one recovers the mild discrepancy previously observed between these two kinds of 
%fits~\cite{DescotesGenon:2001tn}, whereas the larger uncertainty of the $S$-$P$ fit covers both solutions.
%
%
%
\begin{table}[t]
{\small 
%\begin{center}
\begin{tabular}{|c||c|c|c||c|c|c|}
\hline & \multicolumn{3}{|c||}{With isospin-breaking corrections} &
\multicolumn{3}{|c|}{Without isospin-breaking corrections} \\
\hline
& {$S$-$P$}
& {Extended}
& {Scalar} 
& {$S$-$P$}
& {Extended}
& {Scalar}
\\
\hline
$a_0^0$ & $0.221\pm 0.018$ 
              & $0.232\pm 0.009$ 
              & $0.226 \pm 0.007$
              & $0.247\pm 0.014$
              & $0.247\pm 0.008$
               & $0.242\pm$ 0.006\\
$a_0^2$  & $-0.0453\pm 0.0106$ 
              & $-0.0383\pm 0.0040$ 
              & $-0.0431\pm 0.0019$
              & $-0.0357\pm 0.0096$
              & $-0.0349\pm 0.0038$
               & $-0.0396\pm 0.0015$ \\
$\rho_{a_0^0,a_0^2}$ & 0.964
            & 0.881
            & 0.914
            & 0.945
            & 0.842
            & 0.855\\
$\theta_0$ &    $(82.3\pm 3.4)^\circ$
&     $(82.3\pm 3.4)^\circ$
&    $82.3^\circ$
&     $(82.3\pm 3.4)^\circ$
&     $(82.3\pm 3.4)^\circ$
&   $82.3^\circ$
\\
$\theta_1$ &    $(108.9\pm 2)^\circ$
&    $(108.9\pm 2)^\circ$
&    $108.9^\circ$
&    $(108.9\pm 2)^\circ$
&    $(108.9\pm 2)^\circ$
&    $108.9^\circ$
\\
$\chi^2/N$ & 7.6/6
   & 16.6/16
   & 7.8/8
   & 7.2/6
   & 15.7/16
   & 7.3/8
   \\
\hline
 $\alpha$  & $1.043\pm 0.548$ 
              & $1.340\pm 0.231$ 
              & $1.179\pm 0.123$
              & $1.637\pm 0.472$
              & $1.672\pm 0.208$
               & $1.458\pm 0.098$ \\
 $\beta$     & $1.124\pm 0.053$ 
              & $1.088\pm 0.020$ 
              & $1.116\pm 0.007$
              & $1.103\pm 0.055$
              & $1.098\pm 0.021$
               & $1.128\pm 0.008$ \\
 $\rho_{\alpha\beta}$ & 0.47
        & 0.31
        & 0.02
        & 0.47
        & 0.32
        & 0.00\\
 $\lambda_1 \cdot 10^3$  
              & $-3.56\pm 0.68$
              & $-3.80\pm 0.58$ 
              & $-3.89\pm 0.10$ 
              & $-3.79\pm 0.68$
              & $-3.78\pm 0.57$
               & $-3.74\pm 0.11$
               \\
 $\lambda_2 \cdot 10^3$
               & $9.08\pm 0.28$
              & $8.94\pm 0.10$ 
              & $9.14\pm 0.04$ 
              & $9.02\pm 0.23$
              & $9.02\pm 0.11$
               & $9.21\pm 0.42$ \\

 $\lambda_3 \cdot 10^4$ 
              & $2.38\pm 0.18$ 
              & $2.30\pm 0.14$ 
              & $2.32\pm 0.04$
              & $2.34\pm 0.18$
              & $2.34\pm 0.14$
               & $2.41\pm 3.67$ \\
 $\lambda_4 \cdot 10^4 $
              & $-1.46\pm 0.10$ 
              & $-1.39\pm 0.04$ 
              & $-1.45\pm 0.02$
              & $-1.41\pm 0.10$
              & $-1.40\pm 0.04$
               & $-1.46\pm 0.02$ \\
\hline
$\bar\ell_3$
& $3.15\pm 9.9$ 
              & $-10.2\pm 5.7$ 
              & $-2.7\pm 6.6$
              & $-39.9\pm 20.3$
              & $-43.5\pm 19.1$
               & $-19.6\pm 7.8$ \\
$\bar\ell_4$ 
& 5.3$\pm 0.8$ 
              & $4.4\pm 0.6$ 
              & $5.1\pm 0.3$
              & $5.2\pm 0.8$
              & $5.2\pm 0.7$
               & $6.0\pm 0.4$ \\
$X(2)$        
              & $0.88\pm 0.05$ 
              & $0.80\pm 0.06$
              & $0.82\pm 0.02$
              & $0.72\pm 0.05$
              & $0.71\pm 0.05$ 
               & $0.75\pm 0.03$ \\

$Z(2)$         & $0.87\pm 0.03$ 
              & $0.89\pm 0.02$ 
              & $0.86\pm 0.01$
              & $0.87\pm 0.02$
              & $0.87\pm 0.02$
               & $0.85\pm 0.01$ \\
\hline
\end{tabular}
\caption{Scattering lengths, subthreshold parameters and chiral low-energy constants for the different fits considered, 
with and without the isospin-breaking correction $\delta_{\rm IB} (s ;  a_0^2  , a_0^2  )$.}\label{tab:fitresults}
%\end{center}
}
\end{table}

Once the scattering lengths in the isospin limit have been determined,
we can test $N_f=2$ $\chi$PT by comparing the dispersive and chiral descriptions of the low-energy $\pi\pi$ 
amplitude in the isospin limit. First, the solutions of the Roy equations are used to reconstruct the $\pi\pi$ 
amplitude in the unphysical (subthreshold) region where $\chi$PT should converge particularly well. 
 As explained in Refs.~\cite{Stern:1993rg,Knecht:1995tr} and recalled in Ref.~\cite{DescotesGenon:2001tn}, in 
 the isospin limit, one can describe the $\pi\pi$ amplitude in  terms of only six parameters ($\alpha,\beta,\lambda_1,\lambda_2,\lambda_3,\lambda_4$)
up to and including terms of order $(E/\Lambda_H)^6$ in the low-energy
expansion. These subthreshold parameters yield the $N_f=2$ chiral low-energy constants 
$\bar{\ell}_3,\bar{\ell}_4$, or equivalently the two-flavour 
quark condensate and pion decay constant measured in physical units
\begin{equation}
X(2)=\frac{2m\Sigma(2)}{F_\pi^2M_\pi^2}\,, \qquad Z(2)=\frac{F^2(2)}{F_\pi^2}\,,
\qquad  \Sigma(2)=-\lim_{m_u,m_d\to 0} \langle 0|\bar{u}u|0\rangle\,, \qquad F(2)=\lim_{m_u,m_d\to 0} F_\pi\,,
\end{equation}
by matching the chiral expansions to the subthreshold parameters $\alpha,\beta$. These expansions
are expected to exhibit a good convergence since they involve the $\pi\pi$ scattering amplitude far from  singularities. The corresponding values 
of the subthreshold parameters and of the chiral low-energy constants are gathered in Tab.~\ref{tab:fitresults}.
For comparison, we also show the results obtained
without including the isospin corrections.
One should emphasize that the minor difference in $a_0^2$ between the three fits once isospin-breaking corrections are included 
is sufficient to yield significant differences in the estimate of the $N_f=2$ chiral order parameters and low-energy constants.
We may also point out that, although $X(2)$ and $Z(2)$ are defined in the two-flavour chiral expansion,
their deviations from unity exceed the level usually expected for quark-mass corrections in the $N_f=2$ framework.

\section{Summary - Conclusion}

The high-precision data for $\delta_S (s) - \delta_P (s)$ obtained by the NA48/2 experiment
require that isospin-breaking corrections be taken into account. Since the ultimate goal is to 
extract the values of $a_0^0$ and $a_0^2$, the $\pi\pi$ scattering
lengths in the isospin limit, the corrections should not be computed at fixed values
of the scattering lengths, but should be parameterized in terms of them.
 
We have shown that general properties (analyticity, unitarity, crossing, chiral counting)
provide the necessary tools to do this in a model independent way.
The phases of the two-loop form factors can be computed analytically, and the
isospin-breaking correction $\delta_{\rm IB} (s ;  a_0^2  , a_0^2  )$ can be obtained as a
function of the scattering lengths in the isospin limit. We have thus extended the analysis
of IB correction in Ref. \cite{Colangelo:2008sm} in two respects: by going to two loops
in the low-energy expansion, and by keeping the scattering lengths as free parameters.

We have redone the fit to the NA48/2 data using our determination of $\delta_{\rm IB} (s ;  a_0^2  , a_0^2  )$. 
The results we obtain are compatible with those published by NA48/2 within errors.

%%%%%%%%%%%%%%%%%%%%%%%%%%%%%%%%%%%%%%%%%%%%%%%%
%% BACKMATTER
%%%%%%%%%%%%%%%%%%%%%%%%%%%%%%%%%%%%%%%%%%%%%%%%

\begin{theacknowledgments}
  One of us (MK) would like to thank the organizers of the {\it XIth Conference 
  on Quark Confinement and the Hadron Spectrum} for the invitation
  to present our work, and for creating a very pleasant atmosphere
  during this meeting. We thank B. Bloch-Devaux from the NA48/2 Collaboration
  for stimulating and informative discussions.
\end{theacknowledgments}



\bibliographystyle{aipproc}   % if natbib is available

%%%\bibliography{sample}

%%%\IfFileExists{\jobname.bbl}{}
%%% {\typeout{}
%%%  \typeout{******************************************}
%%%  \typeout{** Please run "bibtex \jobname" to optain}
%%%  \typeout{** the bibliography and then re-run LaTeX}
%%%  \typeout{** twice to fix the references!}
%%%  \typeout{******************************************}
%%%  \typeout{}
%%% }

\end{document}